\documentclass[conference]{IEEEtran}
\IEEEoverridecommandlockouts

\usepackage{cite}
\usepackage{amsmath,amssymb,amsfonts}
\usepackage{algorithmic}
\usepackage{graphicx}
\usepackage{url}
\usepackage{textcomp}
\usepackage{xcolor}
\def\BibTeX{{\rm B\kern-.05em{\sc i\kern-.025em b}\kern-.08em
    T\kern-.1667em\lower.7ex\hbox{E}\kern-.125emX}}
\begin{document}

\title{Bridging the Modality Gap in Long-Form Clinical Audio: A Comparative Study of Lightweight and Heavyweight End-to-End SOAP Generation}

\author{
\IEEEauthorblockN{Ziyu Zhang, Mingchen Shao, Wenjie Tian, Tianlun Zuo, Longhao Li, Lei Xie}
\IEEEauthorblockA{Audio, Speech and Language Processing Group (ASLP@NPU), Northwestern Polytechnical University, China\\
ziyu\_zhang@mail.nwpu.edu.cn, lxie@nwpu.edu.cn}
}

\maketitle


\begin{abstract}
Automating clinical documentation from long-form doctor-patient conversations remains challenging for modern audio-language models. While cascaded ASR systems perform well, end-to-end (E2E) models often struggle with information loss and hallucinations on extended audio. For the BeTraC 2026 challenge, the ASLP team presents a fully E2E multimodal system that generates structured SOAP notes directly from audio, bypassing intermediate transcripts. We constructed a 1.41-million-sample multi-task corpus and applied a multi-stage pipeline: domain pre-training, supervised fine-tuning, and reward optimization. Evaluating the architecture under both Lightweight (3B) and Heavyweight (30B) constraints reveals that each training stage progressively enhances performance. Furthermore, scaling to 30B parameters substantially boosts concept extraction and summarization quality. Ultimately, our E2E systems consistently outperform representative cascaded ASR+LLM baselines, proving the efficacy of direct multimodal optimization for clinical documentation.
\end{abstract}

\vspace{-1ex}
\section{Introduction}
Recent advances in Large Audio Language Models (LALMs) and foundational speech representation~\cite{hubert, seamlessm4t,ltu} have improved short-context audio processing, yet reasoning over long-form audio exceeding five minutes remains a formidable challenge.~\cite{audiolm, speecht5, wav2vec2, conformer} A critical application of this is automating clinical documentation, such as generating structured SOAP (Subjective, Objective, Assessment, and Plan) notes directly from lengthy doctor-patient conversations~\cite{labrak2026generating, clinical_burnout, medpalm, biobert, clinicalbert, chatdoctor}. To push the boundaries of end-to-end (E2E) multimodal understanding, the Beyond Transcription Challenge (BeTraC) at IEEE SLT 2026 tasks models with directly generating clinical notes from raw medical audio, bypassing traditional ASR-to-text cascaded pipelines~\cite{berard_e2e}.  In this paper, the ASLP team presents a robust E2E multimodal system for the BeTraC challenge. To investigate the impact of parameter scale on mitigating long-audio hallucinations, we participated in both the Lightweight and Heavyweight tracks, providing a controlled comparative study. Following recent paradigms in instruction tuning and alignment \cite{instructgpt, dpo}, we constructed a comprehensive multi-task instruction-tuning corpus and proposed a four-stage training paradigm. First, full-parameter pre-training establishes foundational cross-modal alignment. Second, Supervised Fine-Tuning (SFT) enforces adherence to SOAP structural constraints. Third, Generative Reward Policy Optimization (GRPO)~\cite{shao2024deepseekmath} directly targets clinical accuracy using custom reward functions for concept extraction, ROUGE scores, formatting, and length control. Finally, Online Preference Distillation (OPD)~\cite{tunstall2023zephyr} is applied to further refine generation quality and suppress repetitive generation loops.To conclude, our main contributions are as follows:
\begin{itemize}
    \item E2E Clinical Generation Pipeline: We propose a fully multimodal architecture for clinical SOAP generation, utilizing a four-stage training pipelineto bridge the long-audio reasoning gap.
    \item Customized Optimization: We design a specialized multi-task learning strategy and a four-dimensional GRPO reward function that significantly boosts medical concept recall and reduces hallucinations.
    \item Dual-Track Comparative Insights: We present a comprehensive empirical comparison between lightweight and heavyweight models, revealing critical insights into parameter capacity limits and fact-retention in long-form E2E speech understanding.
\end{itemize}

\section{Dataset Construction}
\subsection{Dataset Composition}
Our unified training corpus aggregates over 1.41 million valid samples, totaling approximately 8,650 hours of audio, as summarized in Table~\ref{tab:dataset_summary}. The data is meticulously sampled and merged from ten distinct sources: 

\textbf{Core Clinical SOAP Generation:} The foundational dataset is the official BeTraC Synth-DoPaCo training set~\cite{labrak2026generating}, comprising 7,200 long-form synthetic doctor-patient conversations (mean duration of 9 minutes, max 47 minutes) paired with complete SOAP notes. To supplement this, we incorporated real-world clinical audio dialogues and large-scale TTS-concatenated medical dialogue datasets. These core SOAP tasks account for roughly 70\% of the training weight. 

\textbf{Long-Audio Temporal Grounding:} To enhance the model's ability to maintain context over extended sequences and prevent temporal drift, we integrated 8,098 general long-audio temporal positioning and dense audio captioning (DAC/TAC/TAG) samples, alongside 2,679 long-audio ASR samples (9-13 minutes per sample) from the MLC-SLM challenge~\cite{mu2026summary}. 

\textbf{Medical Subtasks and NLP:} We augmented the corpus with real medical ASR datasets and specific SOAP subtasks such as Chief Complaint (CC) extraction and medical QA. Furthermore, seven classic biomedical NLP datasets were converted into an instruction-tuning format to bolster the model's medical terminology comprehension.

\vspace{-1ex}
\begin{table}[htbp]
\caption{Summary of the Unified Multi-Task Dataset for Full-Parameter Pre-training}
\vspace{-4ex}
\begin{center}
\resizebox{\columnwidth}{!}{%
\begin{tabular}{llr}
\hline
\textbf{Task Category} & \textbf{Dataset Source} & \textbf{Duration} \\
\hline
\multicolumn{3}{l}{\textbf{\textit{I. General Long-Form Audio}}} \\
\hline
\hspace{3mm} Long-Form ASR & MLC-SLM Challenge & 350.9 h \\
\hspace{3mm} Dense Audio Captioning & General Web Audio & 1,797.9 h \\
\hspace{3mm} Temporal Grounding & DAC/TAC/TAG & $\sim$1,865 h \\
\hline
\multicolumn{3}{l}{\textbf{\textit{II. Medical Speech Recognition}}} \\
\hline
\hspace{3mm} Conversational ASR & YFYeung Medical ~\footnote{\url{https://huggingface.co/datasets/yfyeung/medical}} & $\sim$48 h \\
\hspace{3mm} Utterance-Level ASR & Eka, Hani89~\footnote{\url{https://huggingface.co/datasets/Hani89/medical_asr_recording_dataset}}, MultiMed\_ST~\cite{mo2024multimed} & $\sim$63 h \\
\hspace{3mm} Synthetic Medical ASR & United-Syn-Med & $\sim$1,785 h \\
\hline
\multicolumn{3}{l}{\textbf{\textit{III. Clinical Audio Comprehension}}} \\
\hline
\hspace{3mm} Medical Term Extraction & Derived Keyword Sets & --- \\
\hspace{3mm} Multiple-Choice QA & MedMosaic QA~\cite{rajgarhia2026medmosaic} & 47.9 h \\
\hline
\multicolumn{3}{l}{\textbf{\textit{IV. SOAP Note Generation (Target Task)}}} \\
\hline
\hspace{3mm} Sectional Extraction (CC) & PaulMooney ASR & 7.8 h \\
\hspace{3mm} \textbf{Full SOAP Generation} & \textbf{Synth-DoPaCo (BeTraC)} & \textbf{1,146.3 h} \\
\hspace{3mm} Full SOAP Generation & YFYeung Medical & 152.8 h \\
\hspace{3mm} Full SOAP Generation & MedConv~\cite{qi2025medconv} \& SOAP-Summary (TTS) & 445.7 h \\
\hline
\textbf{\textit{V. Text-Only Medical NLP}} & BC5CDR~\cite{li2016biocreative}, PubMed~\cite{canese2013pubmed}, DDI~\cite{herrero2013ddi}, etc. & Text-only \\
\hline
\textbf{Total (Union)} & \textbf{---} & \textbf{$\approx$7,760 h} \\
\hline
\end{tabular}%
}
\label{tab:dataset_summary}
\end{center}
\end{table}
\vspace{-3ex}


\subsection{Multi-Task Formulation}
To facilitate joint training across these diverse datasets, we standardized all samples into the ms-swift messages format~\cite{zhao2025swift}, unifying them under 9 distinct instruction prompt templates. These templates cover full SOAP generation, section-specific extraction, ASR transcription, medical QA, keyword extraction, and dense audio description. By mapping all tasks into a unified \texttt{<audio>} + text instruction $\rightarrow$ response paradigm, the model learns to dynamically allocate attention across long audio sequences based on the specified clinical or temporal task constraints.

\begin{table*}[htbp]
\caption{Summary of Best Evaluation Results per Training Stage on the Dev/Test Set.}
\vspace{-2ex}
\begin{center}
\resizebox{\textwidth}{!}{%
\begin{tabular}{llcccccc}
\hline
\textbf{Model} & \textbf{Training Stage} & \textbf{Concept F1} & \textbf{C-Precision} & \textbf{C-Recall} & \textbf{ROUGE-2} & \textbf{ROUGE-3}  & \textbf{Words} \\
\hline
\multicolumn{8}{l}{\textbf{\textit{Lightweight Track ($<$6B)}}} \\
\hline
Qwen2.5-Omni-3B~\cite{hui2024qwen2} & Baseline (Zero-Shot) & 0.2604 & 0.2891 & 0.2450 & 0.0920 & 0.0344  & 380 \\
Qwen2.5-Omni-3B & Best SFT  & 0.3872 & 0.3814 & 0.4093 & 0.2155 & 0.1325  & 410 \\
\textbf{Qwen2.5-Omni-3B} & \textbf{Best GRPO } & \textbf{0.4082} & 0.4078 &  0.4224 & 0.2323 & 0.1418 & 372 \\
\hline
\multicolumn{8}{l}{\textbf{\textit{Heavyweight Track ($<$36B)}}} \\
\hline
Qwen3-Omni-30B~\cite{yang2025qwen3} & Baseline (Zero-Shot) & 0.1879 & 0.1879 & 0.1964 & 0.0558 & 0.0175  & 351 \\
Qwen3-Omni-30B & Best Pre-train  & 0.5129 & 0.4940 & 0.5450 & 0.3410 & 0.2274 & 375 \\
Qwen3-Omni-30B & Best SFT  & 0.5136 & 0.4981 & 0.5413 & 0.3386 & 0.2259  & 367\\
\textbf{Qwen3-Omni-30B} & \textbf{Best GRPO} & \textbf{0.5167} & 0.4732 & 0.5511  & 0.4093 & 0.2323 & 371 \\
\hline
\multicolumn{8}{l}{\textbf{\textit{Reference Systems}}} \\
\hline
Whisper-Large-v3~\cite{radford2023robust} + Qwen3 & Cascaded System & 0.2860 & 0.2964 & 0.2838 & 0.1174 & 0.0425  & 261 \\
Qwen3-ASR-1.7B ~\cite{shi2026qwen3} + Qwen3 & Cascaded System & 0.2772 & 0.2881 & 0.2741 & 0.1092 & 0.0374  & 256 \\
\hline
\vspace{-6ex}
\end{tabular}%
}
\label{tab:best_experiments}
\end{center}
\end{table*}

\section{System Architectures}
\label{sec:architecture}

To satisfy the BeTraC requirement of ``no intermediate transcription'', we adopt a unified end-to-end multimodal architecture instead of conventional cascaded ASR+LLM pipelines. To investigate the impact of model scale on long-form clinical reasoning, we evaluate both Lightweight and Heavyweight configurations under the same framework.

\subsection{End-to-End Multimodal Architecture}

The system directly maps patient-clinician conversations to structured SOAP notes in a single forward pass.~\cite{tang2024salmonn} A frozen audio encoder first extracts acoustic representations, which are projected into the LLM embedding space through a frozen audio-to-LLM aligner.~\cite{chu2023qwen} A fully fine-tuned dense LLM decoder then jointly attends to the acoustic features and textual instruction to autoregressively generate the final SOAP note, without producing any intermediate transcript.

\subsection{Lightweight and Heavyweight Configurations}

The Lightweight system is built upon Qwen2.5-Omni-3B. To satisfy the parameter constraint, unused modules (e.g., the vision tower and speech-generation heads) are removed, resulting in an active dense model with approximately 3B inference parameters.

The Heavyweight system adopts Qwen3-Omni-30B while intentionally preserving the same end-to-end, single-pass architecture. Although the challenge permits external tools such as RAG~\cite{lewis2020retrieval} or CoT~\cite{wei2022chain}, we deliberately avoid them to isolate the effect of model scaling under an identical inference pipeline.

\subsection{Inference Pipeline}

At inference, the input audio is concatenated with a fixed instruction prompt defining the required SOAP structure, as shown in Fig.~\ref{fig:prompt_template}. The same prompt is used during both training and inference, enabling zero-shot generation without additional system prompts or in-context demonstrations.

\begin{figure}[htbp]
\centering
\resizebox{0.95\columnwidth}{!}{%
\begin{tabular}{p{1.05\columnwidth}}
\hline
\textbf{Input Prompt (Role: User)} \\
\hline
\texttt{<audio> This is a patient-doctor conversation recording. Generate a clinical SOAP note following the structure below.} \\
\texttt{**1. Subjective**} \\
\texttt{- Chief Complaint (CC)} \\
\texttt{- History of Present Illness (HPI)} \\
\texttt{- Review of Systems (ROS), grouped by body system} \\
\texttt{**2. Objective**} \\
\texttt{- Physical Examination \& Tests/Results} \\
\texttt{**3. Assessment and Plan**} \\
\texttt{- Group findings as numbered problems: Problem \#1...} \\
\hline
\end{tabular}
}
\caption{Instruction prompt used for end-to-end SOAP note generation.}
\vspace{-2ex}
\label{fig:prompt_template}
\end{figure}

To improve decoding stability on long recordings, we employ deterministic greedy decoding ($T=0$, $p=1.0$) with a repetition penalty of 1.15 and a maximum generation length of 2048 tokens. Finally, a minimal post-processing script only extracts the \texttt{dialog\_id} and preserves the model output verbatim, ensuring that all evaluations reflect the model's native generation quality.

\section{Training Methodology}
\label{sec:training}

We adopt a multi-stage training strategy consisting of supervised fine-tuning (SFT) followed by Group Relative Policy Optimization (GRPO).

\textbf{SFT.} After domain pre-training, the model is fully fine-tuned to generate structured SOAP notes directly from audio inputs. This stage establishes the basic audio-to-text alignment and learns the required clinical document format.


\textbf{GRPO.} To further improve factual consistency and reduce hallucinations, we optimize the model with GRPO. Aligning with the recent paradigm of Reinforcement Learning from AI Feedback (RLAIF) \cite{rlaif}, we design a weighted reward function comprising four components:

\begin{itemize}
    \item \textbf{Concept F1:} Rewards accurate extraction of clinical concepts while penalizing unsupported medical claims.
    \item \textbf{ROUGE-2 \cite{rouge}:} Alongside other standard automated metrics \cite{bleu} and LLM-as-a-judge frameworks \cite{llm_judge}, this component explicitly encourages lexical overlap with the reference SOAP notes.
    
    \item \textbf{SOAP Format:} Ensures compliance with the required document structure.
    \item \textbf{Length Penalty:} Suppresses excessively long or repetitive generations.
\end{itemize}

\textbf{OPD.} Finally, to further align the model's generation quality and mitigate structural failures, we apply Online Preference Distillation. Specifically, we employ a Chain-of-Thought (CoT) distillation strategy where the model is jointly trained on prompts with and without CoT reasoning trajectories. By treating the robust CoT outputs as preferred responses, we distill complex clinical reasoning capabilities directly into the standard, single-pass generation policy. This ensures that the model maintains high factual accuracy at inference—strictly complying with the "no intermediate transcription" rule—while also acting as a strong regularizer to suppress the infinite repetition loops occasionally triggered by ultra-long audio sequences.

\vspace{-1ex}
\section{Experiments and Results}
\label{sec:experiments}

\subsection{Experimental Setup}
Models were developed using the SWIFT framework \cite{zhao2025swift} to support efficient distributed training. For the Heavyweight pipeline (Qwen3-Omni-30B), training was conducted on a 4-node cluster comprising 32 NVIDIA H20 GPUs with a global batch size of 16 and a learning rate of $5e^{-6}$. Conversely, the Lightweight pipeline (Qwen2.5-Omni-3B) was distributed across 8 GPUs with an effective batch size of 16 and a learning rate of $1e^{-6}$. During inference, deterministic greedy decoding (Temperature $T=0, p=1.0$) was rigidly applied across both tracks to ensure reproducibility. To mitigate structural loop hallucinations on ultra-long audio, a repetition penalty of $1.15$ was strictly enforced.

\subsection{Main Results and Comparative Analysis}
Table~\ref{tab:best_experiments} shows the best performance achieved at each training stage for both the Lightweight and Heavyweight tracks, together with two cascaded ASR+LLM reference systems.

For the Lightweight track, progressive supervised fine-tuning and reinforcement learning consistently improved the end-to-end model. Starting from a zero-shot Concept F1 of 0.2604, SFT raised the score to 0.3872, while the subsequent GRPO optimization further improved it to 0.4082. Similar upward trends were observed on ROUGE-2 and ROUGE-3, demonstrating that reinforcement learning provides crucial complementary gains beyond supervised instruction tuning.

The Heavyweight track exhibited substantially stronger performance throughout training. The zero-shot model achieved only 0.1879 Concept F1, indicating that the pretrained multimodal checkpoint was not natively aligned with the clinical SOAP generation task. After domain-specific pre-training on the full corpus, the Concept F1 increased dramatically to 0.5129. Subsequent SFT produced a comparable performance (0.5136), while GRPO further pushed the final Concept F1 to an impressive 0.5167. Although the final gain from GRPO was relatively modest for the large model, it consistently produced the best overall Concept F1 and ROUGE scores.

Comparing the two tracks, scaling the backbone from 3B to 30B yielded a substantial improvement under the exact same end-to-end architecture and training pipeline. The Heavyweight configuration outperformed the Lightweight model by an absolute margin of 0.108 in Concept F1 (0.5167 vs. 0.4082). This suggests that increased model capacity significantly enhances long-context latent reasoning and clinical information retention from extended conversational audio.

Finally, both end-to-end systems substantially outperformed the cascaded ASR+LLM baselines. The best Lightweight model exceeded the Whisper-Large-v3 and Qwen3-ASR based pipelines by an absolute Concept F1 margin of over 0.122, while the Heavyweight model widened this margin to a remarkable 0.230. These results empirically demonstrate that directly optimizing an end-to-end multimodal model is considerably more effective than relying on intermediate speech transcription for long-form clinical summarization.

\vspace{-1ex}
\section{Conclusion}
\label{sec:conclusion}
We presented ASLP’s end-to-end multimodal systems for the BeTraC 2026 challenge. Using multi-task instruction tuning and clinical reward-based GRPO, our systems directly generate structured SOAP notes from long-form audio. Results show that model scaling improves factual accuracy and reduces hallucinations, while lightweight models remain limited by representation capacity. Future work will explore memory mechanisms and acoustic token compression for more efficient clinical reasoning.
\bibliographystyle{IEEEtran}
\bibliography{mybib}

\end{document}